\documentclass[12pt,a4paper,english]{article}
\pdfoutput=1
\usepackage{mathptmx}
\usepackage{helvet}
\usepackage{courier}
\usepackage[T1]{fontenc}
\usepackage[latin9]{inputenc}
\usepackage{babel}

\usepackage{amsmath}
\usepackage{graphicx}
\usepackage[unicode=true, pdfusetitle,
 bookmarks=true,bookmarksnumbered=false,bookmarksopen=false,
 breaklinks=false,pdfborder={0 0 1},backref=false,colorlinks=false]
 {hyperref}
\hypersetup{
 pdfkeywords={Quantum mechanics, Bohmian interpretation, Bohmian mechanics, particle trajectories, Bell inequality, CHSH inequality, Stern-Gerlach experiment}}
 
\begin{document}

\title{Bohmian particle trajectories contradict quantum mechanics}

\author{\noindent Michael Zirpel}
\maketitle
\begin{abstract}
\noindent The Bohmian interpretation of quantum mechanics adds particle
trajectories to the wave function and ensures that the probability
distribution of the particle positions agrees with quantum mechanics
at any time. This is not sufficient to avoid contradictions with quantum
mechanics. There are correlations between particle positions at different
times which cannot be reproduced with real particle trajectories.
A simple rearrangement of an experimental test of the Bell-CHSH inequality
demonstrates this. 
\end{abstract}

\section*{Introduction}

\noindent The Bohmian interpretation \cite{BIBITEM.1e5a728a-99ca-421a-b369-c9f0d896b57f}
of quantum mechanics (also called Bohmian mechanics \cite{BIBITEM.e72b6d2f-67f6-4ba7-80cf-67f4c3384f5f})
supplements the quantum mechanical wave function with particle trajectories.
These trajectories are thought to be \emph{real} in the following
sense:
\begin{enumerate}
\item Each particle moves along a trajectory and has at any time $t\hspace{0.222222em}$a
definite position $\boldsymbol{Q}(t)$ on its trajectory. 
\item The result of a position measurement at time $t$ is the particle
position $\boldsymbol{Q}(t)$. 
\end{enumerate}
\noindent In an ensemble of equally prepared particles with the same
wave function $\mathrm{\psi}$, each particle moves along a randomly
chosen trajectory. The \emph{quantum equilibrium hypothesis} and \emph{Bohm's
law of motion} \cite{BIBITEM.e72b6d2f-67f6-4ba7-80cf-67f4c3384f5f}
guarantee that the quantum mechanical probability distribution of
the particle positions

\[
\mathrm{\rho}(\boldsymbol{r},t)=|\mathrm{\psi}(\boldsymbol{r},t)|^{2}\]

\noindent is maintained for all times $t$ by the Bohmian theory.\\

\noindent However, the real particle movement along trajectories entails
correlations between particle positions at different times. As R.
F. Streater noticed in \cite{BIBITEM.5153097f-a0e2-48bb-afc0-665bc720beae}
(p. 108) this must contradict quantum mechanics in certain cases because
particle positions at different times are represented by non-commuting
operators in the Heisenberg picture. Accordingly a Bell-type inequality
could be established between certain functions of the particle position
at different times, which is fulfilled by trajectories but violated
by quantum mechanics.\\

\noindent A simple way to demonstrate this contradiction is to change
the standard experimental setting for the test of the CHSH inequality
\cite{BIBITEM.45aea027-c721-4e2d-9000-59d2199af6dc}, \cite{BIBITEM.eab81f81-5c8b-49f2-84fb-bcafc194398a}
so that all measured quantities are represented by different particle
beams in one setting. The Bohmian particle trajectories follow some
of these beams so that all quantities will have definite values in
each run of the experiment. This implies compliance with the CHSH
inequality, which is however violated by quantum mechanics in certain
cases.

\section*{CHSH test with Stern-Gerlach apparatuses}

\noindent The basic experimental setting was already discussed by
D. Bohm in his 1951 text book \cite{BIBITEM.5e49c65e-5e70-46cf-a148-ce38477d9d0d}
in the context of the EPR paradox. The source $S$ emits pairs of
neutral spin $\frac{1}{2}$ particles in opposite directions, where
Stern-Gerlach apparatuses are used to measure spin components: The
particle beams are split by inhomogeneous magnetic fields $M_{A}$,
$M_{B}$ in different directions. Detection screens $S_{A},$$S_{B}$
complete the measurement. If a particle is detected in the 'upper'
\footnote{The terms 'upper', 'lower' are related to the diagram and not to the
real direction.%
} half of the screen the corresponding observables $A$ and $B$, respectively
have the value $+1$, if the particle is detected in the 'lower' half,
the corresponding observables have the value $-1$. \\

\noindent \begin{center}
\includegraphics[scale=0.75]{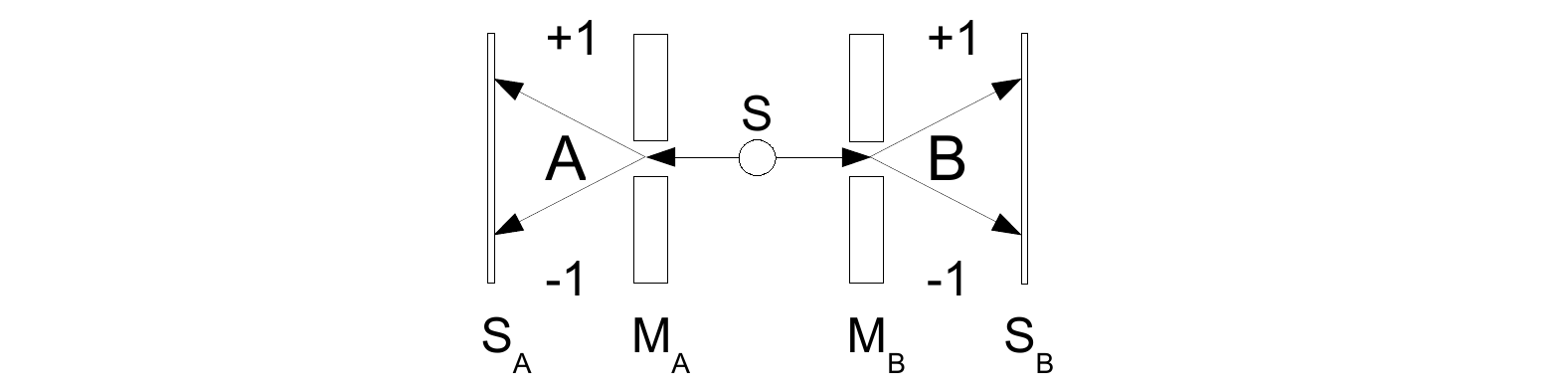}
\par\end{center}

\noindent \begin{center}
Figure 1: CHSH test experiment with Stern-Gerlach apparatuses
\par\end{center}

\noindent In a CHSH test experiment the magnetic fields are switched
between two different directions on each side so that in total four
observables $A,A^{\prime}$ and $B,B^{\prime}$ are measured. With
appropriate directions of the magnetic fields and a suited initial
state of the particle pair the expectation value of $A\hspace{0.222222em}B+A\hspace{0.222222em}B^{\prime}+A^{\prime}\hspace{0.222222em}B-A^{\prime}\hspace{0.222222em}B^{\prime}$
will be

\noindent \begin{equation}
\left|\left\langle A\hspace{0.222222em}B+A\hspace{0.222222em}B^{\prime}+A^{\prime}\hspace{0.222222em}B-A^{\prime}\hspace{0.222222em}B^{\prime}\right\rangle \right|\approx2\sqrt{2}\label{EQUATION.255fc06e-75bb-44e1-9f8e-f2479e030af5}\end{equation}

\noindent which exceeds the bound $2$ given by the CHSH inequality
\cite{BIBITEM.77bc8a26-dfdb-44de-b876-ad45b3f89dd9}.

\section*{Four observables in one apparatus}

\noindent If the detection screens $S_{A},$$S_{B}$ are removed,
the partial beams can be recombined coherently using other magnetic
fields so that the original spin state is restored. This is possible
in theory, because the evolution without measurement is unitary and
therefore reversible as already has been noticed by D. Bohm in his
textbook \cite{BIBITEM.5e49c65e-5e70-46cf-a148-ce38477d9d0d} p.605
(the technical difficulties of the experimental realization are discussed
in \cite{BIBITEM.d89221d6-a9a1-44ea-a39e-ec98fcbb0762}). After the
coherent recombination another pair of magnetic fields $M_{A'}$,
$M_{B'}$ can be used to split the beam again to measure the observables
$A^{\prime},B^{\prime}$. \\

\noindent \begin{center}
\includegraphics[scale=0.75]{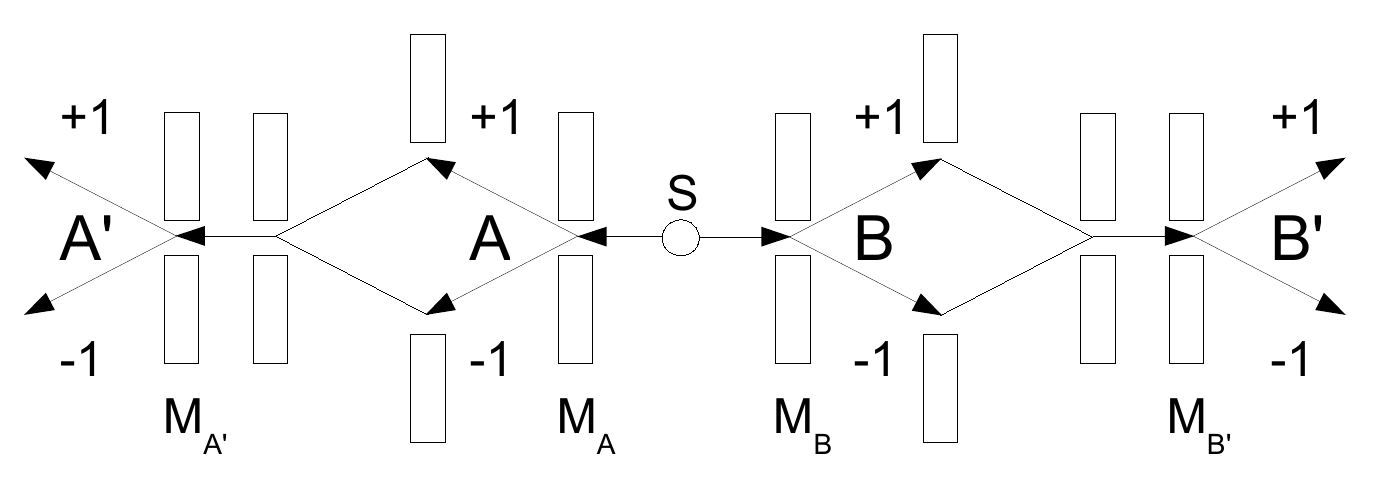}
\par\end{center}

\noindent \begin{center}
Figure 2: Four observables in one apparatus
\par\end{center}

\noindent In this new setting all four observables $A,A^{\prime},B,B^{\prime}$
are represented by different partial beams in one apparatus. They
are functions of the particle positions, which are in one-to-one correspondence
with some particle spin components.\\
By the placement of the detection screens the experimenter decides
which pair of observables is measured. The expectation value of $A\hspace{0.222222em}B+A\hspace{0.222222em}B^{\prime}+A^{\prime}\hspace{0.222222em}B-A^{\prime}\hspace{0.222222em}B^{\prime}$
will be the same as above \eqref{EQUATION.255fc06e-75bb-44e1-9f8e-f2479e030af5}.

\section*{\noindent Bohmian trajectories fulfill the CHSH inequality}

\noindent Bohmian trajectories in a Stern-Gerlach apparatus with packet-like
wave function are analyzed in \cite{BIBITEM.b66c891e-5c7f-4c75-a527-4eadc5a57c01},
\cite{BIBITEM.ec36c0cd-cdff-452a-96a3-578f1378fa72}. Nearly all trajectories
follow the particle beams. Where the beam splits, a part of the trajectory
set follows one partial beam, the rest follows the other partial beam. 

\noindent In a single run of the experiment a pair of Bohmian particles
follows a randomly chosen pair of trajectories, one in direction $A$,
one in direction $B$. The particle moving in direction $A$ moves
along its trajectory either through half-plane $A=+1$ or half-plane
$A=-1$ and afterwards either through half-plane $A^{\prime}=+1$
or half-plane $A^{\prime}=-1$. The same is true for the $B$ particle
and the quantities $B,B^{\prime}$.

\noindent \begin{center}
\includegraphics[scale=0.75]{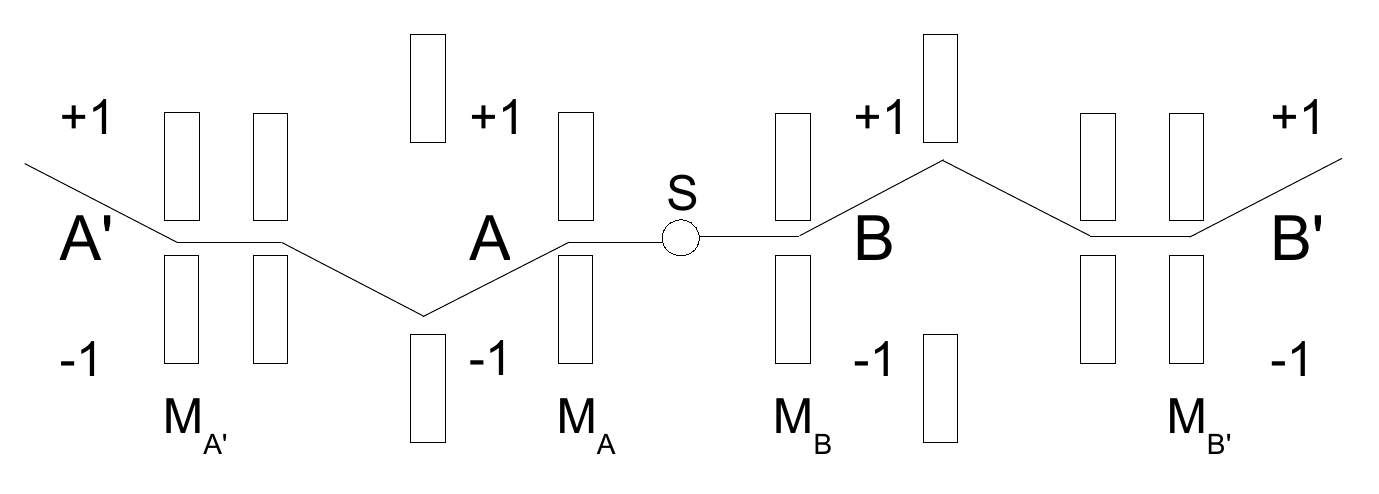}
\par\end{center}

\noindent \begin{center}
Figure 3: A Bohmian trajectory pair with $A=-1,A'=1$ and $B=1,B'=1$ 
\par\end{center}

\noindent So for each particle pair emitted by the source all four
quantities $A,A^{\prime},B,B^{\prime}$ will have a \emph{definite}
value. According to proposition 2) these definite values will be measured,
when detection screens are placed into the appropriate positions.
Definite values, however, always fulfill the CHSH inequality: Let
$a_{k},a_{k}^{\prime},b_{k},b_{k}^{\prime}\in\left\{ -1,+1\right\} $
be the values of the four quantities $A,A^{\prime},B,B^{\prime}$
in the $k$-th run of the experiment. A simple valuation table shows
that always

\[
a_{k}b_{k}+a_{k}b_{k}^{\prime}+a_{k}^{\prime}b_{k}-a_{k}^{\prime}b_{k}^{\prime}=\pm2\]

\noindent So the mean value in $n$ runs of the experiment is bounded
by $2$ for all $n>0$

\begin{equation}
\left|\frac{1}{n}\underset{k=1}{\overset{n}{\sum}}\left(a_{k}b_{k}+a_{k}b_{k}^{\prime}+a_{k}^{\prime}b_{k}-a_{k}^{\prime}b_{k}^{\prime}\right)\right|\le2\label{EQUATION.aabb8321-dd62-4f98-a690-8ab61cca3056}\end{equation}

\noindent The quantum mechanical expectation value \eqref{EQUATION.255fc06e-75bb-44e1-9f8e-f2479e030af5}
violates this inequality.

\section*{Conclusion}

\noindent This demonstrates a contradiction between a quantum mechanical
probability statement about particle positions and the consequences
of the Bohmian theory. As long as propositions 1) and 2) are thought
to be valid, this contradiction cannot be removed. So we must conclude:
\begin{verse}
\emph{Bohmian particle trajectories contradict quantum mechanics}.
\end{verse}

\section*{Acknowledgements}

Thanks to Prof. Bernhard Lauth and Gerhard Zoubek for many stimulating
discussions and constant encouragement.

\end{document}